\documentclass[11pt,a4paper,superscriptaddress,prb,amsmath,amssymb,aps]{revtex4}

\usepackage{graphicx}
\usepackage{bm}

\begin{document}
\title{Hannay angle and geometric phase shifts under adiabatic parameter changes in classical dissipative systems.} 

\author{N.A. Sinitsyn}
\affiliation{Center for Nonlinear Studies and Computer, Computational and Statistical Sciences Division, Los Alamos National Laboratory, Los Alamos, NM 87545 USA}

\author{J. Ohkubo}
\affiliation{Institute for Solid State Physics, University of Tokyo, Kashiwanoha 5-1-5, Kashiwa, Chiba 277-8581, Japan}


\begin{abstract}
In Phys. Rev. Lett. {\bf 66}, 847 (1991), T. B. Kepler and M. L. Kagan derived a geometric
phase shift in dissipative limit cycle evolution. This effect
was considered as an extension of the geometric phase in classical mechanics. 
We show that the converse is also true, namely,  this geometric phase can be identified with 
the classical mechanical Hannay angle in an extended phase space. Our results suggest that this phase
can be generalized to a stochastic evolution with an additional noise term in evolution equations.
\end{abstract}

\date{\today}

\maketitle 
The Berry phase in quantum mechanics \cite{berry-84} appeared as a unifying concept. It provided similar mathematical background to
seemingly very different quantum mechanical phenomena.
After its discovery, various geometric phases were found even beyond
quantum mechanics domain, for example, in classical mechanics \cite{hannay-85}, hydrodynamics \cite{wilczek-88}, 
dissipative kinetics \cite{kagan-91, landsberg-92}, and stochastic processes \cite{sinitsyn-07epl, sinitsyn-07prl, ohkubo-08, shi}.
It was possible to relate some of these phases to each other. For example, the Hannay angle can be derived in the
classical limit of the quantum mechanical Berry phase \cite{berry-book}. Relations and hierarchy among other geometric phases remain not well
understood. In this Letter we partly fill this gap, and demonstrate the relation of the geometric phase in dissipative limit cycle evolution
to the classical mechanical Hannay angle. Our approach is similar to the one employed in \cite{sinitsyn-07prl} to introduce 
geometric phases in stochastic processes, which suggests that further connections among various geometric phases can be found.
 
 In \cite{kagan-91} Kagan {\it et al.} considered a dissipative system that evolves 
to a limit cycle so that after fast relaxation processes the only one angle 
degree of freedom $\phi(t)$ is relevant
and evolves according to 
\begin{equation}
\frac{d\phi}{dt} = \Omega(\phi,{\bf \mu}), 
\label{ev1}
\end{equation}
where ${\bf \mu}$ is the vector of slowly time-dependent parameters and $\Omega$ is the instantaneous rotation frequency.
 Kagan {\it et al.} introduced another angle variable
\begin{equation}
\theta(\phi,{\bf \mu}) = \int_0^{\phi}\frac{\omega({{\bf \mu}}) }{\Omega(\phi',{\bf \mu})} d\phi' ,
\label{theta1}
\end{equation}   
where
\begin{equation}
\omega({\bf \mu}) = \left( \int_0^{2\pi}\frac{1}{\Omega(\phi,{\bf \mu})} \frac{d\phi}{2\pi} \right)^{-1},
\end{equation}
and showed that under the adiabatic cyclic evolution of ${\bf \mu}$ during time $T$,  the phase (\ref{theta1}) becomes the sum of dynamic and geometric parts, i.e.
\begin{equation}
\theta = \theta_{dyn}+\theta_{geom},
\end{equation} 
where 
\begin{equation}
\theta_{dyn}=\int_0^Tdt \omega ({\bf \mu}(t)),
\end{equation}
 and
\begin{equation}
\theta_{geom}=\oint {\bf A \cdot} d{\bf \mu}, 
\quad {\bf A}=\int_0^{2\pi}\frac{d\phi}{2\pi}  \frac{\omega ({\bf \mu}(t))}{\Omega(\phi,{\bf \mu})} \partial_{{\bf \mu}} \theta(\phi,{\bf \mu}).
\label{theta_geom}
\end{equation}
Authors of \cite{kagan-91} argued that  the Hannay angle in classical mechanics is merely a special case of this phase.
Below we show that in some sense the opposite is also true, namely that the geometric phase (\ref{theta_geom}) follows from  canonical 
equations of motion and is identified with the  Hannay 
angle \cite{berry-book}.
 Lets introduce the variable $\Lambda$, which we assume to be canonically conjugated to $\phi$ with the Hamiltonian
\begin{equation}
H(\Lambda,\phi)=\Lambda \Omega(\phi,{\bf \mu}).
\end{equation}
The phase evolution (\ref{ev1}) then follows from equation 
\begin{equation}
\frac{d\phi}{dt}= \frac{\partial H}{\partial \Lambda}.
\label{ev3}
\end{equation}
 In case of time-independent ${\bf \mu}$, its conjugated equation 
\begin{equation}
\frac{d\Lambda}{dt}=-\frac{\partial H}{\partial \phi}
\label{ev 4}
\end{equation}
 has a solution
\begin{equation}
\Lambda = \frac{E({\bf \mu})}{\Omega (\phi,{\bf \mu})},
\end{equation}
where $E$ is the energy. 
The adiabatically conserved quantity is the action defined by
\begin{equation}
I=\frac{1}{2\pi}\int_0^{2\pi}\Lambda(\phi,\mu) d\phi=\frac{1}{2\pi}\int_0^{2\pi}\frac{E({\bf \mu})}{\Omega (\phi,{\bf \mu})}d\phi,
\label{I1}
\end{equation}
from which follows that
\begin{equation}
E({\bf \mu}) =  I \omega({\bf \mu}),
\label{E1}
\end{equation}
and 
\begin{equation}
\Lambda=\Lambda(I,\phi)=I \frac{\omega({\bf \mu})}{\Omega(\phi, {\bf \mu})}.
\end{equation}
Expression for the canonically conjugated to $I$ angle variable $\theta$ reads
\begin{equation}
\theta=\frac{\partial}{\partial I}\left (\int_0^{\phi}\Lambda (I,\phi')d\phi'\right) = \int_0^{\phi}\frac{\omega({\bf \mu})}{\Omega(\phi', {\bf \mu})}d\phi'.
\label{theta_a}
\end{equation}
Comparing (\ref{theta1}) and (\ref{theta_a}) we find that the angle variable $\theta$ introduced in \cite{kagan-91} is just a canonical angle variable in the 
model with the Hamiltonian $H(\Lambda,\phi)$. This, in fact, justifies the choice of variables made in \cite{kagan-91}.
After adiabatic evolution in the parameter
space the angle variable becomes a sum of the dynamic part and the Hannay angle \cite{berry-book}: 
\begin{equation}
\theta=\theta_{dyn} + \theta_H,
\end{equation}
where
\begin{equation}
\theta_{dyn}=\frac{\partial}{\partial I} \int_0^T dt E({\bf \mu}(t)) = \int_0^T \omega({\bf \mu}(t)) dt,
\end{equation}
\begin{equation}
\theta_{H}=-\frac{\partial}{\partial I} \oint d{\bf \mu} 
\langle \Lambda \left(I,\phi(\theta,{\bf \mu})\right) \partial_{{\bf \mu}} \phi(\theta,{\bf \mu}) \rangle_{\theta}=\oint {\bf A \cdot}d{\bf \mu},
\label{din}
\end{equation}
and the averaging is over one fast cycle of $\theta$ angle. The connection ${\bf A}$ explicitly reads
\begin{equation}
{\bf A}=-\int_0^{2\pi} \frac{d\theta}{2\pi} \frac{\omega({\bf \mu})}{\Omega(\phi(\theta,{\bf \mu}),{\bf \mu})}  \partial_{{\bf \mu}} \phi(\theta,{\bf \mu}).
\label{con2}
\end{equation}
The last step is to show that connections in (\ref{con2}) and (\ref{theta_geom}) are the same.
For this notice that 
\begin{equation}
\frac{d\phi(\theta,{\bf \mu})}{dt}= \Omega=\frac{\partial \phi}{\partial \theta} \frac{d\theta}{dt} +
\frac{\partial \phi}{\partial {\bf \mu}} \frac{d{\bf \mu}}{dt}.
\label{int1}
\end{equation}
and that 
\begin{equation}
\frac{d\theta}{dt} = \omega({\bf \mu}) +\frac{\partial \theta}{\partial {\bf \mu}} \frac{d{\bf \mu}}{dt}, \quad 
\frac{\partial \phi}{\partial \theta}=\frac{\Omega}{\omega},  
\label{int2}
\end{equation}
which lead to
\begin{equation}
\partial_{{\bf \mu}}\phi = - \frac{\partial \phi}{\partial \theta} \partial_{{\bf \mu}}\theta.
\label{int3}
\end{equation}
Substituting this into (\ref{con2}) and 
switching to integration over $\phi$ one recovers Eq. (\ref{theta_geom}), which completes our proof that the phase in (\ref{theta_geom}) can be identified
with a Hannay angle.

{\em In conclusion,} we established relation between classical mechanical Hannay angle and the geometric phase in dissipative limit cycle
evolution.  
 Hannay angle interpretation of the phase in \cite{kagan-91} relates it also to geometric phases in stochastic kinetics,
introduced in \cite{sinitsyn-07prl} by similar variable doubling technique, which may be practically interesting. The doubling
of variables can be used to promote not only dissipative but also
stochastic equations to the Hamiltonian evolution \cite{kamenev-04}. Thus it is possible to derive Hamiltonian formulation for Eq. (\ref{ev1}) with an 
additional noise term. However, the physical meaning of the resulting Hannay angle remains an open problem.

\begin{acknowledgments} 
{\it  This work was funded in part by DOE under Contract No.
  DE-AC52-06NA25396.} 
\end{acknowledgments}


\begin{thebibliography}{99}
                                   

                                   



\bibitem{berry-84} M. Berry {\em Proc. R. Soc. Lond. A} {\bf 392}, 45 (1984).
\bibitem{hannay-85} J. H. Hannay  {\em J. Phys. A.} {\bf 18} 221 (1985).

\bibitem{wilczek-88}
A.\ Shapere and F.\ Wilczek, {\em Phys. Rev. Lett.} {\bf 58}, 2051 (1987);
 A.\ Shapere and F.\ Wilczek, {\em J.\ Fluid Mech.}
  {\bf 198}, 557 (1988).

 
\bibitem{kagan-91} M.\ L.\ Kagan, T.\ B.\ Kepler and I.\ R.\ Epstein,  {\em Nature} {\bf 349}, 506 (1991); T.\ B.\ Kepler and  M.\ L.\ Kagan,
  Phys. Rev. Lett. 66, 847 (1991).
\bibitem{landsberg-92} A. S. Landsberg, {\em Phys. Rev. Lett.} {\bf 69}, 865 (1992).

\bibitem{sinitsyn-07epl} N.\ A.\ Sinitsyn and I.\ Nemenman, {\em EPL}
  {\bf 77}, 58001 (2007); preprint q-bio/0612018.
\bibitem{sinitsyn-07prl} N.\ A.\ Sinitsyn, and I.\ Nemenman, {\em Phys. Rev. Lett.} {\bf 99}, 220408 (2007).
\bibitem{ohkubo-08} Jun Ohkubo, {\em J. Stat. Mech.} P02011 (2008).
\bibitem{shi} Y.\ Shi and Q.\ Niu, {\em Europhys.\ Lett.} {\bf 59}, 324
  (2002).
\bibitem{berry-book} D.\ Chruscinski, and A.\ Jamiolkowski, "Geometric Phases in Classical and Quantum Mechanics", Birkh\"auser, (2004). 
\bibitem{kamenev-04} 
V.\ Elgart and A.\ Kamenev, {\em Phys. Rev. E}
  {\bf 70}, 041106 (2004).







 



\end{thebibliography}
\end{document}